**Title:** JWST molecular mapping and characterization of Enceladus' water plume feeding its torus

**Author list:** G. L. Villanueva[1,*], H. B. Hammel[2], S. N. Milam[1], V. Kofman[1,3], S. Faggi[1,3], C. R. Glein[4], R. Cartwright[5], L. Roth[6], K. P. Hand[7], L. Paganini[8], J. Spencer[9], J. Stansberry[10], B. Holler[10], N. Rowe-Gurney[1,11], S. Protopapa[9], G. Strazzulla[12], G. Liuzzi[13], G. Cruz-Mermy[14], M. El Moutamid[15], M. Hedman[16], K. Denny[16]

**Affiliations:**

(1) NASA Goddard Space Flight Center, Greenbelt MD 20771, USA

* Corresponding author: geronimo.villanueva@nasa.gov

(2) Association of Universities for Research in Astronomy, Washington DC 20004, USA

(3) American University, Washington, DC 20016, USA

(4) Southwest Research Institute, San Antonio TX 78238, USA

(5) Carl Sagan Center for Research at the SETI Institute, Mountain View CA 94043, USA

(6) KTH Royal Institute of Technology, Stockholm 104 50, Sweden

(7) Jet Propulsion Laboratory, Pasadena CA 91109, USA

(8) NASA Headquarters, Washington DC 20546, USA

(9) Southwest Research Institute, Boulder CO 80302, USA

(10) Space Telescope Science Institute, Baltimore MD 21218, USA

(11) Department of Astronomy & CRESST II, University of Maryland, College Park MD 20742

(12) INAF-Osservatorio Astrofisico di Catania, 95123 Catania, Italy

(13) Università degli Studi della Basilicata, 85100 Potenza, Italy

(14) Universite Paris-Sarclay, 91190 Gif-sur-Yvette, France

(15) Cornell University, Ithaca NY 14853, USA

(16) University of Idaho, Moscow ID 83844, USA

**Abstract**

Enceladus is a prime target in the search for life in our solar system, having an active plume likely connected to a large liquid water subsurface ocean. Using the sensitive NIRSpec instrument onboard JWST, we searched for organic compounds and characterized the plume's composition and structure. The observations directly sample the fluorescence emissions of $H_2O$ and reveal an extraordinarily extensive plume (up to 10,000 km or 40 Enceladus radii) at cryogenic temperatures (25 K) embedded in a large bath of emission originating from Enceladus' torus. Intriguingly, the observed outgassing rate (300 kg/s) is similar to that derived from close-up observations with Cassini 15 years ago, and the torus density is consistent with previous spatially unresolved measurements with Herschel 13 years ago, suggesting that the vigor of gas eruption from Enceladus has been relatively stable over decadal timescales. This level of activity is sufficient to maintain a derived column density of $4.5 \times 10^{17}$ $m^{-2}$ for the embedding equatorial torus, and establishes Enceladus as the prime source of water across the Saturnian system. We performed searches for several non-water gases ($CO_2$, CO, $CH_4$, $C_2H_6$, $CH_3OH$), but none were identified in the spectra. On the surface of the trailing hemisphere, we observe strong $H_2O$ ice features, including its crystalline form, yet we do not recover $CO_2$, CO nor $NH_3$ ice signatures from these observations. As we prepare to send new spacecraft into the outer solar system, these observations demonstrate the unique ability of JWST in providing critical support to the exploration of distant icy bodies and cryovolcanic plumes.

**Main text**

**Introduction**

Enceladus is likely the largest source of water within the Saturnian system[1], with $H_2O$ and other materials jetted into Saturn orbit by localized geological activity[2,3]. Early hints of geological activity on Enceladus were provided by Voyager and telescopic observations in the 1980s and 1990s, finding a close association between Saturn's E ring and Enceladus' orbit[4–8]. In 2005, multiple instruments onboard the Cassini spacecraft discovered a plume of gases (predominately water vapor) and ice grains emerging from fissures in the south polar region of Enceladus[9–14], while a torus of water along Enceladus' orbit was most recently observed via sub-millimeter spectroscopy with the Herschel Observatory[15]. The Cassini measurements of the plume gas were made using *in situ* mass spectrometry along specific flyby trajectories[16,17] and via stellar occultation in the inner region of the plume (<200 km)[9,18]. In contrast, the sub-millimeter measurements of the torus were not spatially resolved, but they indicated the presence of $H_2O$ gas widely throughout the Saturnian system. While the plume's flux of icy grains varies on multiple timescales[19], the variations in the vapor flux are much less well understood, together with how these affect the structure and evolution of the torus. By analyzing the molecular emissions across large distances from Enceladus with JWST[20], we were able to map the distribution of outgassed water, compare the level of activity to that determined by Cassini measurements, and establish a direct connection of the plume to the extended cloud of material beyond the plume that likely accumulated over multiple orbits.

On November 9th 2022 UT, we observed Enceladus' trailing hemisphere with JWST as part of the Solar System Guaranteed-Time-Observations (GTO) program 1250. The JWST/NIRSpec observations were made with the Integral Field Unit (IFU)[21], delivering a data-cube across three high-resolving power gratings (G140H, G235H, G395H), two detectors per grating (NRS1, NRS2), with a uniform spaxel size of 0.1″x0.1″ across a 3″x3″ field-of-view (FOV) – Enceladus was 0.07″ in diameter at the time of the observations. To minimize saturation, we employed the NRSRAPID readout and short integration times per frame, totaling 215-270 seconds of integration per grating. The data were processed employing the latest version of the JWST Science Calibration Pipeline (v1.9), and we developed ad-hoc algorithms to analyze the frames, combine dithering images, and clean bad pixels (scripts publicly available at github.com/nasapsg). In Figure 1, we present flux-calibrated spectra for the integrated signal across the Enceladus disk and molecular residuals along several regions of the extended plume.

**Results, surface ices:**

The Enceladus disk spectrum as observed with JWST is dominated by water ($H_2O$) ice, including its crystalline form (features at 1.65 and 3.1 µm), with the main features similar to those identified using Cassini's Visible and Infrared Mapping Spectrometer (VIMS)[2] and from ground-based observations[22,23]. Carbon dioxide ($CO_2$) ice features were previously identified near 2.7 µm and 4.3 µm on the surface of Enceladus and primarily at southern polar latitudes (80-90°S) from VIMS[2,24], but in the JWST spectrum we do not identify either of these $CO_2$ ice bands. The non-detection of $CO_2$ on Enceladus with JWST is likely due to the observing geometry at the time of the observations (centered at 16°N, 270°W), which did not sufficiently sample Enceladus' southern polar regions, further establishing that $CO_2$ ice is likely confined to the south polar terrain and probably replenished by plume $CO_2$[2,25]. We do observe a potential signature near 4.5 µm, which could be related to CN compounds, yet this identification is only tentative from these data. More information about this and other possible ice identifications is given in the methods section (M2).

We also searched for ammonia ($NH_3$) ice, which can strongly depress the freezing point of water and could indicate a sub-surface liquid water reservoir that feeds the plume[17], but we did not identify it in the spectrum (either amorphous or crystalline). Spectroscopic searches for crystalline $NH_3$ ice in the near infrared have mostly focused on a feature near 2.24 μm, attributed to a combination mode. Ground-based observations of the trailing side of Enceladus detected a subtle band centered near 2.25 μm that was tentatively attributed to $NH_3$ ice, but other observations obtained in 1995 and 1998 did not display this feature[22]. We did notice a weak and broad feature spanning ~2.15 to 2.2 μm that has been attributed to $NH_3$-$H_2O$ mixtures and ammonium ($NH_4$) bearing species on other icy moons[26]. However, this absorption was inconclusive between dithers and not consistent with the 2.24 μm expected location for $NH_3$ ice, nor the previously detected 2.25 μm feature on Enceladus. Additionally, there is no clear evidence of the ~2.0 μm absorption band due to solid ammonia and its hydrates[27] and expected to overlap with the much stronger 2.0 μm water ice band. We also searched for the stronger $NH_3$ ice band at 2.96 μm[28] but the NIRSpec spectra do not show any absorption at this wavelength. Our non-detection of $NH_3$ ice agrees with the non-detection of surface $NH_3$ on a global scale by Cassini VIMS[2].

**Results, plume and torus mapping:**

The uniqueness of JWST for exploring Enceladus is most evident when probing with unparalleled sensitivity the narrow infrared emissions emanating from the plume. Such probing is possible thanks to JWST's very low operational temperatures (~40 K), broad wavelength coverage (0.6 to 28.3 μm), large collecting area (25.4 m$^2$), high spectral resolving power (up to λ/δλ~3000) and advanced infrared instrumentation (e.g., IFU, multi-grating). Infrared gas emissions at these wavelengths are primarily due to solar-pumped fluorescence, so they are particularly weak at these large heliocentric distances. Furthermore, the molecular features are narrow and confined to at most twice the expansion velocity ($v_{exp}$ ~ 540 m/s)[1] requiring high-resolution spectroscopy, which is sensitively done with JWST since we can also probe the strong molecular fundamentals of $H_2O$ and $CO_2$ (not accessible to ground-based observatories due to the lack of atmospheric transparency). Across the Enceladus plume, we detect several strong $H_2O$ molecular emissions (Figure 1) – also at the spaxel level (Figure 2) – revealing a highly localized source oriented southward and extending out to at least 10,000 km (40 $R_E$). We retrieved column densities from the measured line fluxes by employing the publicly available Planetary Spectrum Generator (PSG) tool at psg.gsfc.nasa.gov[29,30], which integrates non LTE (Local Thermodynamic Equilibrium) fluorescence radiative-transfer models[31–33] (see S1). From the relative intensities of the water ro-vibrational lines, we determined that the water molecules are at a rotational temperature of $T_{rot}$ = 25 ± 3 K, and this temperature is consistent across the different regions explored (i.e., local plume, extended plume, background emission) by our measurements (Figure 1). These observed temperatures refer to the rotational excitation temperature, not to the ambient kinetic temperature of $H_2O$ vapor. Due to the large extension of the plume and low local densities, this sparse collisional regime dictates that the molecules are no longer in LTE but in an equilibrium state defined by the insolation rate and the intrinsic probability of absorption or emission of the molecule.

The observed map of water column densities can be relatively well modeled[30] (see Figure 2) when considering an ejecta plume with an outgassing rate of (1.0±0.1)×10$^{28}$ molecules per second (300 kg/s), a 3D ejecta cone half-width of 40° as inferred from these data, and an expansion velocity of 540 m/s (an intermediate value between the ~400 m/s thermal velocity[3], the vertical velocity of ~600 m/s as inferred from near-surface measurements[34], and the ~700 m/s flow velocities inferred from an outgassing model[16]). This observed level of activity is similar to that inferred from measurements[1,18] made in Saturn orbit by Cassini 6-19 years ago ($1\times10^{28}$ molecules per second), although some other analyses suggest large temporal variability in outgassing from Enceladus[16,35]. We also observe a relatively constant background water emission across the whole FOV with an average column density of $(1.7\pm0.1)\times10^{18}$ m$^{-2}$. Such emission most likely comes from water molecules within the torus, which is being observed at the ansa of the orbital path of Enceladus with an inclination of 15.2°. If we assume a constant average density within the torus and zero density outside, and take into account that the torus minor radius or scale height is considerably smaller than its major radius (~237,000 km, centered on Enceladus' orbit), then the equatorial column density cutting perpendicular through the torus center can be estimated to be $N_T \sim 4.5\times10^{17}$ m$^{-2}$ ($1.7\times10^{18}$ m$^{-2}$ · sin 15.2°). Our measurement is remarkably consistent with that inferred from sub-millimeter observations[15] obtained 13 years ago, which indicated an equatorial column density of $4\times10^{17}$ m$^{-2}$ and a torus scale height of $H_T \sim$ 25,000 km.

**Discussion**

As Enceladus orbits rapidly around Saturn with a period of only 1.37 Earth days, the ejected water vapor is spread along and around its orbit, forming a large torus around Saturn. Considering that the photochemical lifetime[36] of water near Saturn is relatively long (~94 days) and combining this value with the derived production rate of the plume, we estimate that up to

$8\times10^{34}$ molecules are available at a given time. Alternatively, and from our derived torus equatorial column density ($N_T$) and the inferred torus scale height ($H_T$), we estimate that a total of $2.5\times10^{34}$ molecules are confined within the torus, equivalent to 32% of the ejected molecules. This would mean that a large fraction of the ejected $H_2O$ molecules (and its OH and O products) are spread beyond the torus and across the Saturnian system. These results generally agree with the findings by Cassidy and Johnson[1], establishing Enceladus as the dominant source of exogenous $H_2O$ / OH / O species in the Saturnian system. In addition to $H_2O$ vapor, we searched for $CO_2$, CO, $CH_4$, $C_2H_6$ and $CH_3OH$ molecular emissions across the plume, but none were detected (see extended Figure 1 and the methods section M1). Upper limits ($3\sigma$) on their abundances are respectively: <1%, <10%, <4%, <6% and <20% relative to water. These limits are within the abundances reported from Cassini INMS measurements[25,37] of the dense plume region of Enceladus ($CO_2$:0.3-0.8%, CO<0.05%, $CH_4$:0.1-0.3%, $C_2H_6$<0.2%, $CH_3OH$<0.01%). UVIS occultations also showed no evidence of CO (<1%) in the inner regions of the plume[18]. Our upper limit on the $CO_2/H_2O$ ratio provides additional support for the idea[38] that extensive $CO_2$ sequestration in Enceladus' rocky core is probably needed to explain why its plume is significantly depleted in $CO_2$ compared with cometary observations[39,40].

**Conclusions:**

These first observations with JWST (only a few minutes of integration time) demonstrate the power of this observatory for sensitively characterizing this ocean world, opening a new window into the exploration of Enceladus' ongoing plume activity while preparing for future missions[41]. More generally, JWST can provide detailed quantitative insights into $H_2O$ vapor-dominated geological and cryovolcanic activity elsewhere in the solar system.

**Acknowledgements**

This work was supported by NASA's Goddard Astrobiology Program, Goddard's Fundamental Laboratory Research (FLaRe), and the Sellers Exoplanet Environments Collaboration (SEEC). This work is based on observations made with the NASA/ESA/CSA James Webb Space Telescope. The data were obtained from the Mikulski Archive for Space Telescopes at the Space Telescope Science Institute, which is operated by the Association of Universities for Research in Astronomy, Inc., under NASA contract NAS 5-03127 for JWST. These observations are associated with program #1250. SNM, HBH and NRG acknowledge support from NASA JWST Interdisciplinary Scientist grant 21-SMDSS21-0013. CRG was supported by Southwest Research Institute Internal Research & Development grant 15-R6248 and NASA Astrobiology Institute grant NNN13D485T. KPH acknowledges support from the NASA Astrobiology Program (award #80NSSC19K1427) and the Europa Lander Pre-Project, managed by the Jet Propulsion Laboratory, California Institute of Technology, under a contract with NASA.

**Data availability**

The data used in this analysis are publicly available at the Space Telescope Science Institute (STScI) JWST archive (https://mast.stsci.edu/), program #1250.

**Code availability**

The retrieval software package used in this study is the Planetary Spectrum Generator, which is free and available online at https://psg.gsfc.nasa.gov[29,30], with the data-reduction scripts available at https://github.com/nasapsg. Figures were made with Matplotlib version 3.2.1[42], available under the Matplotlib license at https://matplotlib.org/.

**Competing Interests Statement**

The authors declare no competing interests.

**Author Contributions Statement**

Villanueva, Hammel, Milam, Hand, Paganini, Spencer, Stansberry and Strazzulla designed the observations and prepared the observational plans. Villanueva, Kofman, Faggi, Cartwright, Stansberry, Holler, Protopapa, Liuzzi, Hedman and Denny analyzed the data, extracted calibrated spectra, produced maps and performed retrievals. Glein, Roth, Rowe-Gurney, Cruz-Mermy, and El-Moutamid assisted with the interpretation of the results, and provided context to related mission and astronomical investigations. All authors contributed to the preparation, writing and edition of the manuscript.

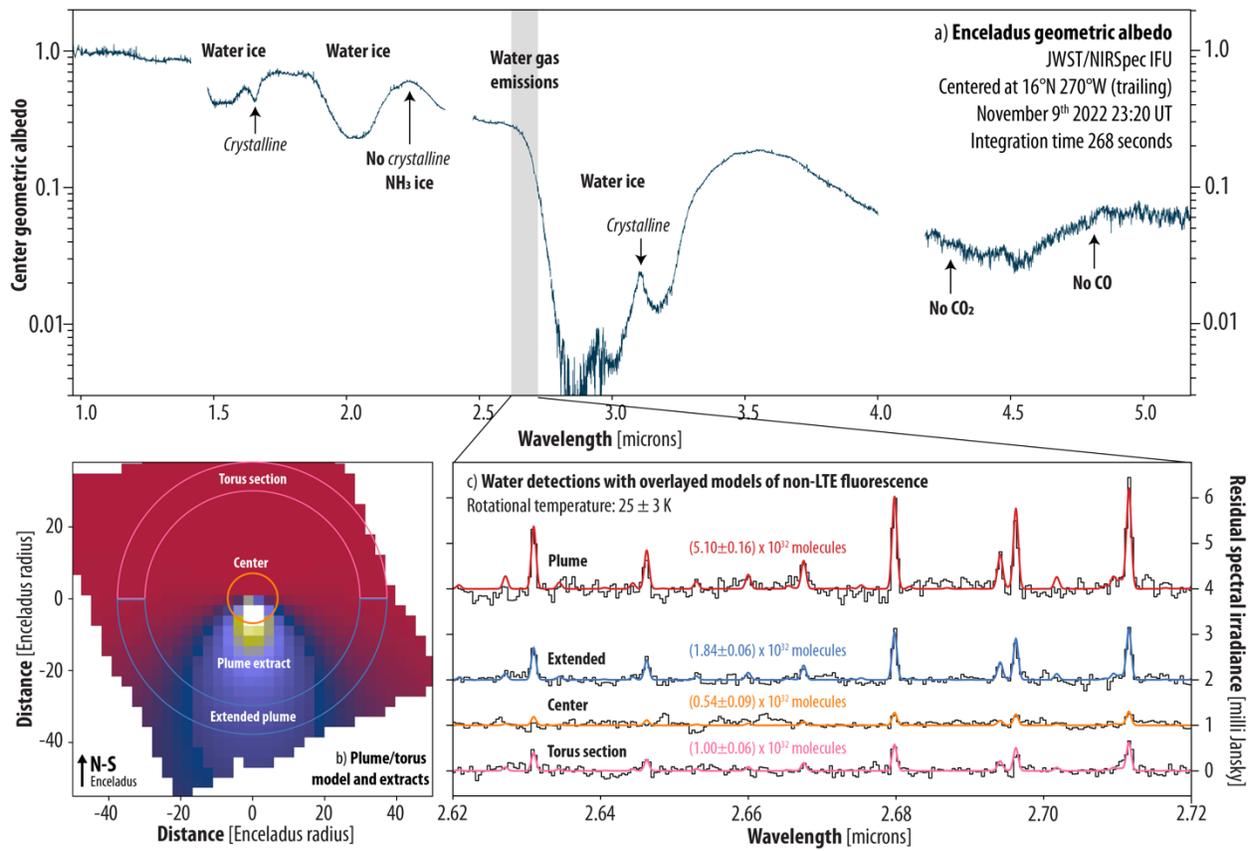

**Figure 1:** Enceladus' surface geometric albedo and detected water vapor emissions. a) Surface geometric albedo of the trailing hemisphere normalized with respect to a reflected solar model[30]. The spectrum shows several strong signatures of $H_2O$ ice, while no absorptions are observed at the expected wavelengths for $CO_2$, CO or $NH_3$ ice. b) Model of the observed water outgassing, in which 4 distinct regions are identified: the center region (orange circle) within 7 Enceladus radii ($R_E$); the inner plume region between 7 and 30 $R_E$; the extended plume region (blue contour) towards the South and between 30 and 38 $R_E$; and the torus background region (pink contour) towards the North and between 30 and 38 $R_E$. c) Data (black lines) and model (colored lines) of the $H_2O$ fluorescence emissions within the four regions of panel b, shifted vertically for clarity. The retrieved number of molecules for each region is also indicated. All models are consistent with a rotational temperature of 25 ±3K.

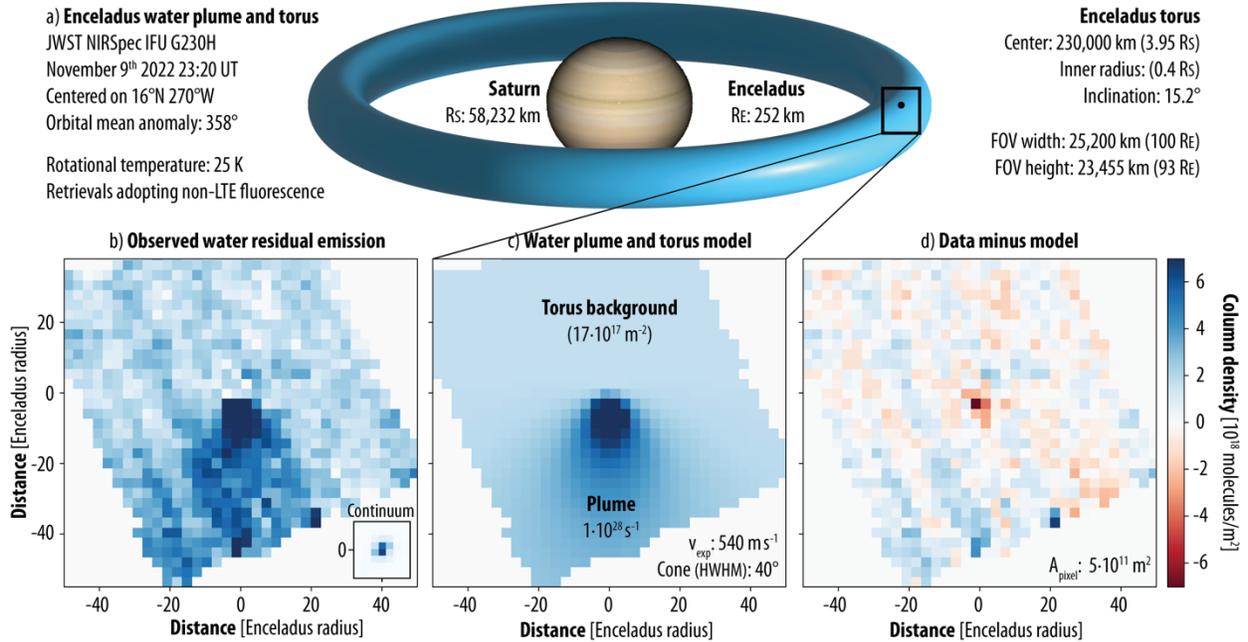

**Figure 2:** Water emission is observed across the whole field of view, revealing an immense water plume emanating from Enceladus which is feeding an extended background torus around Saturn. a) The observations sample the trailing hemisphere of Enceladus and the edge of the torus, where $R_S$ refers to the mean radius of Saturn. b) At each spaxel (0.1"×0.1"), the $H_2O$ column density was retrieved from the observed molecular fluorescence emissions in the 2.62 to 2.72 μm range. Enceladus is 0.07" in diameter (smaller than a spaxel), and the continuum image of the Point-Spread-Function (PSF) is shown in the inset box. Some residual diagonal striping is observed, which we suspect originates from detector effects. c) A model[30] with two components as shown in Figure 1b, consisting of a plume and a torus background emission, reproduce the observations well. d) A residual image was computed by subtracting the outgassing model from the observations, revealing a close fit to the data.

## Methods

### M1. Retrievals of plume molecular species and derivations of upper-limits

In order to search for narrow molecular features, we analyzed the residual spectra which were derived by subtracting a continuum model that included solar Fraunhofer lines from the observed Enceladus spectra. We performed integrations across several regions throughout the environment of Enceladus (see Figure 1) and determined the number of molecules and corresponding column densities. We detect signatures of water vapor in all regions, with the most prominent molecular features and detections across the plume region ($7<\rho<30$ $R_E$, 0.2 to 1.0", 1765 to 7563 km). This region also has a low intrinsic continuum signature from Enceladus' disk which simplifies the removal of the non-gas signatures. We therefore used this region to search for other gases beyond $H_2O$. The fluorescence models are based on non-LTE radiative-transfer modeling[31–33]. Broad non-molecular features were removed by fitting a polynomial function to the continuum shape over the spectral regions presented in Extended Figure 1. Retrievals and the statistical analysis were performed using PSG, in which the retrieval algorithm is based on the Optimal Estimation method [43]. After each iteration of the retrieval algorithm, a new model was constructed, and numerical derivatives were computed for each parameter. This process was repeated until convergence was achieved, and the differences between data and model were minimized. The mean statistical variation of the residual spectra (RMS or chi-square) was used to quantify the uncertainty (sigma) in the retrieved column densities.

### M2. Additional findings regarding the characterization of the icy surface of Enceladus

The trailing hemispheric spectrum of Enceladus measured with NIRSpec does show some interesting features, which could provide additional information regarding the composition and physical properties of Enceladus' surface. Yet these findings are not conclusive at this stage and additional laboratory experiments, observations, analysis, and modeling would be required to further establish the significance of these findings.

**Origin of the water crystalline features:** The presence of the 1.65 μm and 3.11 μm features that we observe in the Enceladus spectrum (Figure 1), clearly testifies to the crystallinity of the $H_2O$ ice[44,45]. Interestingly, laboratory spectra[46] (1.3 to 2.5 μm) of a thin film of crystalline ice deposited at 150 K and cooled down to 16 K show this 1.65 μm crystalline band. On the other hand, crystalline ice formed at higher temperatures does not exhibit this feature, which is also absent in amorphous ice[44]. From the relative intensities of these features and their observed central wavelengths one could obtain constraints on the formation and current temperature of the observed ices, yet this would also need detailed modeling of the other strong nearby features that impact the shape of the nearby continuum.

**Hydrogen peroxide-bearing ice:** We observe a subtle "plateau" at 3.5 μm (see Figure 1), which could be attributed to hydrogen peroxide ($H_2O_2$), yet proper recovery of this feature would require accurate modeling of the nearby strong water bands. It is well known that the icy satellites within the Saturnian and Jovian magnetospheres are subjected to intense fluxes of energetic particles that alter their surficial properties and induce many physical and chemical effects[47,48]. Although modeling how efficiently radiation processes affect Enceladus' surface quantitatively will need to account for the relatively low energetic proton fluxes[49,50], and relatively high particle deposition rates of plume fallout[51] experienced by the Saturnian moon. Among the several effects studied in the laboratory is the formation of $H_2O_2$ evidenced from the appearance of a band at about 3.5 µm in the spectrum of water ice irradiated with energetic particles[52–54]. This 3.5 µm $H_2O_2$ feature has been found on the surface of Europa[55,56], while ultraviolet observations made by the Galileo Ultraviolet Spectrometer suggest that $H_2O_2$ may be present on Ganymede and Callisto as well[57]. Consequently, the possible 3.5 µm feature on Enceladus could be consistent with radiolytic generation of $H_2O_2$ from its $H_2O$ ice rich surface.

**CN compounds:** A possible absorption is observed near 4.5 µm (see Figure 1) and could be associated to CN and/or (iso)cyanate compounds, yet baseline issues for these low flux continuum levels makes this identification only tentative. Radiolytic and photochemical processing of C-containing ices results in the formation of an organic material that, formed at low temperatures, evolves during the heating of the samples and yields an organic refractory material that is stable at room temperature and above[58]. When the original ice contains both C and N atoms, the stable residue exhibits a strong and clear feature centered at about 4.6 µm that effectively reproduces the features observed in some ultra-carbonaceous Antarctic meteorites[59]. That feature is attributed to cyanate and isocyanate bonds and is considered evidence of the energetic processing (by photons, electrons or ions) of the ices. Similarly, a feature centered near 4.57 µm was detected on Callisto and attributed to CN-bearing organics[60]. Such ices on Enceladus could be deposited on the surface by micrometeoritic bombardment or by energetic processing of reduced C- and N-bearing materials. Another possibility is that CN compounds might already be present in Enceladus' plume[37], and could then be deposited over the surface.

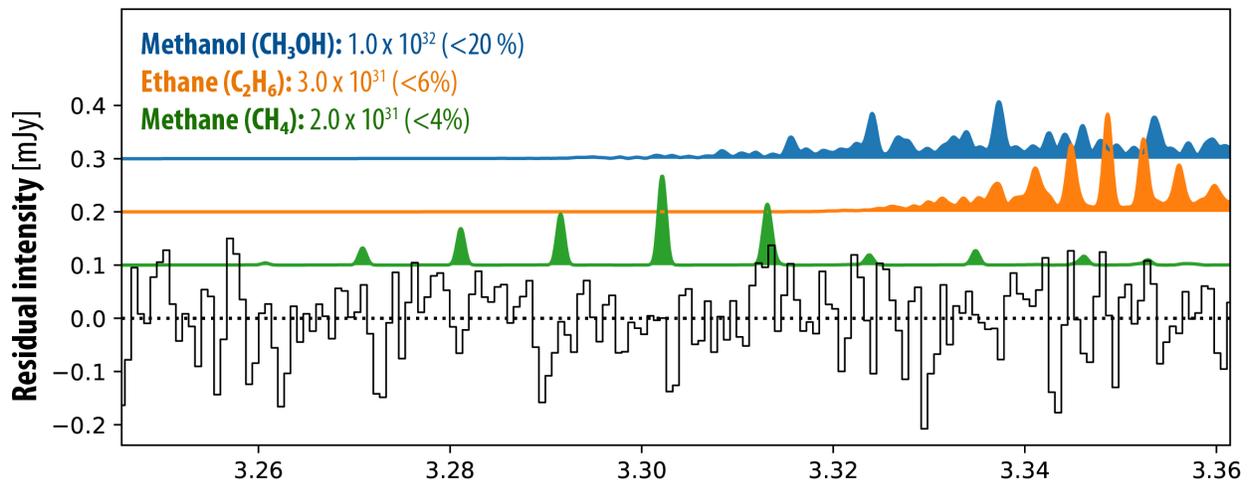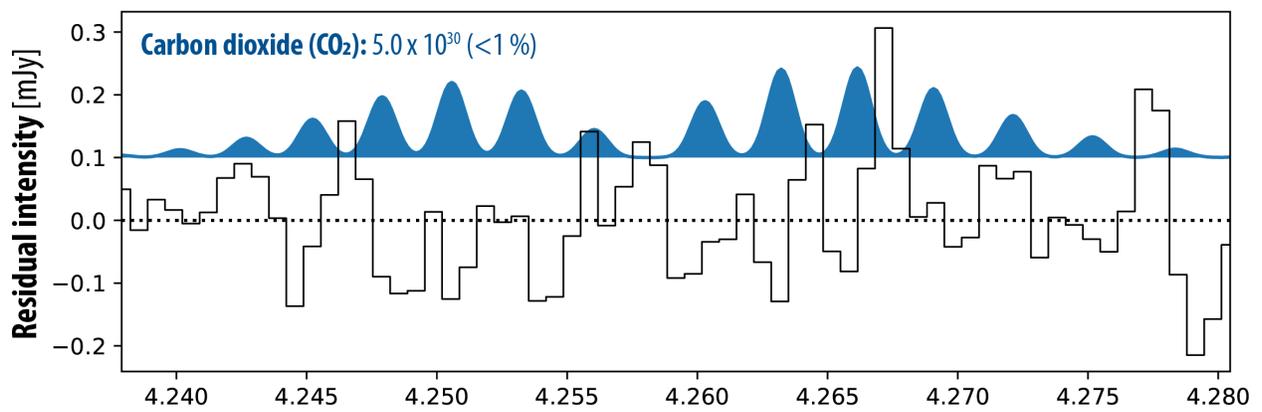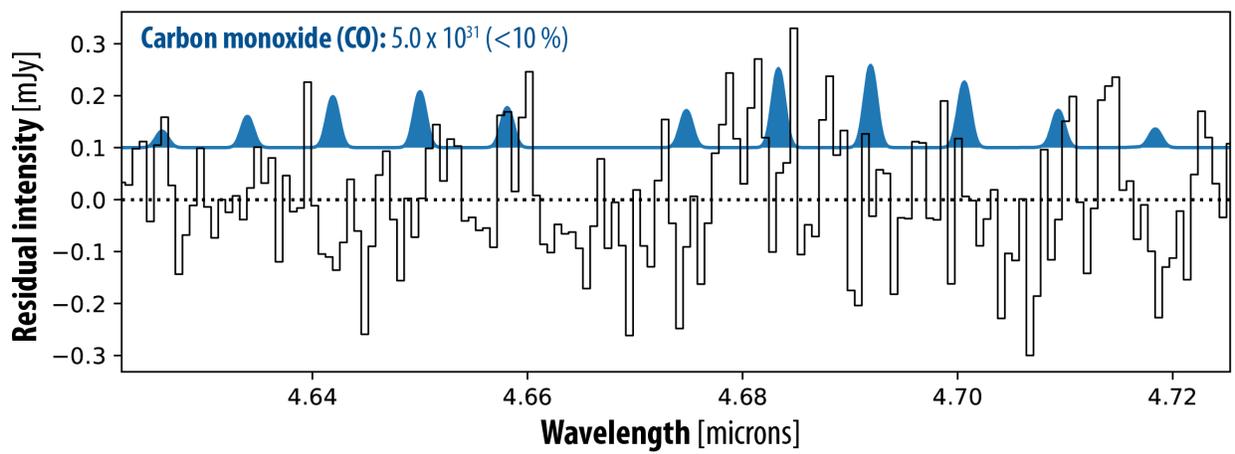

**References**

1. Cassidy, T. A. & Johnson, R. E. Collisional spreading of Enceladus' neutral cloud. *Icarus* **209**, 696–703 (2010).

2. Brown, R. H. *et al.* Composition and Physical Properties of Enceladus' Surface. *Science* **311**, 1425–1428 (2006).

3. Spencer, J. R. *et al.* Cassini Encounters Enceladus: Background and the Discovery of a South Polar Hot Spot. *Science* **311**, 1401–1405 (2006).

4. Baum, W. A. *et al.* Saturn's E ring: I. CCD observations of March 1980. *Icarus* **47**, 84–96 (1981).

5. Haff, P. K., Eviatar, A. & Siscoe, G. L. Ring and plasma: The enigmae of Enceladus. *Icarus* **56**, 426–438 (1983).

6. Smith, B. A. *et al.* A New Look at the Saturn System: The Voyager 2 Images. *Science* **215**, 504–537 (1982).

7. Shemansky, D. E., Matheson, P., Hall, D. T., Hu, H.-Y. & Tripp, T. M. Detection of the hydroxyl radical in the Saturn magnetosphere. *Nature* **363**, 329–331 (1993).

8. Dougherty, M. K., Buratti, B. J., Seidelmann, P. K. & Spencer, J. R. *Enceladus as an Active World: History and Discovery. Enceladus and the Icy Moons of Saturn* 3 (2018). doi:10.2458/azu_uapress_9780816537075-ch001.

9. Hansen, C. J. *et al.* Enceladus' Water Vapor Plume. *Science* **311**, 1422–1425 (2006).

10. Porco, C. C. *et al.* Cassini Observes the Active South Pole of Enceladus. *Science* **311**, 1393–1401 (2006).

11. Waite, J. H. *et al.* Cassini Ion and Neutral Mass Spectrometer: Enceladus Plume Composition and Structure. *Science* **311**, 1419–1422 (2006).

12. Spahn, F. *et al.* Cassini Dust Measurements at Enceladus and Implications for the Origin of the E Ring. *Science* **311**, 1416–1418 (2006).

13. Dougherty, M. K. *et al.* Identification of a Dynamic Atmosphere at Enceladus with the Cassini Magnetometer. *Science* **311**, 1406–1409 (2006).

14. Schenk, P. M., Clark, R. N., Howett, C. J. A., Verbiscer, A. J. & Waite, J. H. *Enceladus and the Icy Moons of Saturn*. (The University of Arizona Press, 2018).

15. Hartogh, P. *et al.* Direct detection of the Enceladus water torus with *Herschel*. *A&A* **532**, L2 (2011).

16. Smith, H. T. *et al.* Enceladus plume variability and the neutral gas densities in Saturn's magnetosphere. *Journal of Geophysical Research: Space Physics* **115**, (2010).

17. Waite, J. H. *et al.* Liquid water on Enceladus from observations of ammonia and 40Ar in the plume. **460**, 487 (2009).

18. Hansen, C. J. *et al.* The composition and structure of Enceladus' plume from the complete set of Cassini UVIS occultation observations. *Icarus* **344**, 113461 (2020).

19. Ingersoll, A. P., Ewald, S. P. & Trumbo, S. K. Time variability of the Enceladus plumes: Orbital periods, decadal periods, and aperiodic change. *Icarus* **344**, 113345 (2020).

20. Gardner, J. P. *et al.* The James Webb Space Telescope. *Space Sci Rev* **123**, 485–606 (2006).

21. Böker, T. *et al.* The Near-Infrared Spectrograph (NIRSpec) on the James Webb Space Telescope - III. Integral-field spectroscopy. *A&A* **661**, A82 (2022).

22. Cruikshank, D. *et al.* A spectroscopic study of the surfaces of Saturn's large satellites: HO ice, tholins, and minor constituents. *Icarus* **175**, 268–283 (2005).

23. Emery, J. P., Burr, D. M., Cruikshank, D. P., Brown, R. H. & Dalton, J. B. Near-infrared (0.8–4.0 m) spectroscopy of Mimas, Enceladus, Tethys, and Rhea. *A&A* **435**, 353–362 (2005).

24. Combe, J.-P. *et al.* Nature, distribution and origin of CO2 on Enceladus. *Icarus* **317**, 491–508 (2019).

25. Waite, J. H. *et al.* Cassini finds molecular hydrogen in the Enceladus plume: Evidence for hydrothermal processes. *Science* **356**, 155–159 (2017).

26. Cartwright, R. J. *et al.* Evidence for Ammonia-bearing Species on the Uranian Satellite Ariel Supports Recent Geologic Activity. *ApJL* **898**, L22 (2020).

27. Zheng, W., Jewitt, D. & Kaiser, R. I. INFRARED SPECTRA OF AMMONIA–WATER ICES. *ApJS* **181**, 53 (2009).

28. Roser, J. E., Ricca, A., Cartwright, R. J., Ore, C. D. & Cruikshank, D. P. The Infrared Complex Refractive Index of Amorphous Ammonia Ice at 40 K (1.43–22.73 μm) and Its Relevance to Outer Solar System Bodies. *Planet. Sci. J.* **2**, 240 (2021).

29. Villanueva, G. L., Smith, M. D., Protopapa, S., Faggi, S. & Mandell, A. M. Planetary Spectrum Generator: An accurate online radiative transfer suite for atmospheres, comets, small bodies and exoplanets. *Journal of Quantitative Spectroscopy and Radiative Transfer* **217**, 86–104 (2018).

30. Villanueva, G. L. *et al. Fundamentals of the Planetary Spectrum Generator*. *Fundamentals of the Planetary Spectrum Generator. 2022 edition of the handbook by G.L. Villanueva et al. ISBN 978-0-578-36143-7* (2022).

31. Villanueva, G. L. *et al.* Water in planetary and cometary atmospheres: H$_2$O/HDO transmittance and fluorescence models. *Journal of Quantitative Spectroscopy and Radiative Transfer* **113**, 202–220 (2012).

32. Villanueva, G. L., Disanti, M. A., Mumma, M. J. & Xu, L.-H. A Quantum Band Model of the ν$_3$ Fundamental of Methanol (CH$_3$OH) and its Application to Fluorescence Spectra of Comets. *The Astrophysical Journal* **747**, 1–11 (2012).

33. Villanueva, G. L., Mumma, M. J. & Magee-Sauer, K. Ethane in planetary and cometary atmospheres: Transmittance and fluorescence models of the ν$_7$ band at 3.3 μm. *Journal of Geophysical Research* **116**, 1–23 (2011).

34. Hansen, C. J. *et al.* Water vapour jets inside the plume of gas leaving Enceladus. *Nature* **456**, 477–479 (2008).

35. Teolis, B. D. *et al.* Enceladus Plume Structure and Time Variability: Comparison of Cassini Observations. *Astrobiology* **17**, 926–940 (2017).

36. Huebner, W. F., Keady, J. J. & Lyon, S. P. Solar photo rates for planetary atmospheres and atmospheric pollutants. *Astrophysics and Space Science (ISSN 0004-640X)* **195**, 1–289 (1992).

37. Postberg, F. *et al. Plume and Surface Composition of Enceladus*. *Enceladus and the Icy Moons of Saturn* 129 (2018). doi:10.2458/azu_uapress_9780816537075-ch007.

38. Glein, C. R. & Waite, J. H. The Carbonate Geochemistry of Enceladus' Ocean. *Geophysical Research Letters* **47**, e2019GL085885 (2020).

39. Pinto, O. H., Womack, M., Fernandez, Y. & Bauer, J. A Survey of CO, CO2, and H2O in Comets and Centaurs. *Planet. Sci. J.* **3**, 247 (2022).

40. Ootsubo, T. *et al.* AKARI Near-infrared Spectroscopic Survey for CO2 in 18 Comets. *The Astrophysical Journal* **752**, 15 (2012).

41. MacKenzie, S. M. *et al.* The Enceladus Orbilander Mission Concept: Balancing Return and Resources in the Search for Life. *Planet. Sci. J.* **2**, 77 (2021).

42. Caswell, T. A. *et al.* matplotlib/matplotlib v3.1.3. (2020) doi:10.5281/zenodo.3633844.

43. Rodgers, C. D. *Inverse Methods for Atmospheric Sounding: Theory and Practice*. (World Scientific, 2000). doi:10.1142/3171.

44. Mastrapa, R. M. *et al.* Optical constants of amorphous and crystalline H2O-ice in the near infrared from 1.1 to 2.6 μm. *Icarus* **197**, 307–320 (2008).

45. Mastrapa, R. M., Sandford, S. A., Roush, T. L., Cruikshank, D. P. & Ore, C. M. D. Optical constants of amorphous and crystalline H2O-ice: 2.5–22 μm (4000–455 cm−1). *ApJ* **701**, 1347 (2009).

46. Leto, G., Gomis, O. & Strazzulla, G. The reflectance spectrum of water ice: Is the 1.65 mu msp peak a good temperature probe?. *Memorie della Societa Astronomica Italiana Supplementi* **6**, 57 (2005).

47. Strazzulla, G. Cosmic ion bombardment of the icy moons of Jupiter. *Nuclear Instruments and Methods in Physics Research Section B: Beam Interactions with Materials and Atoms* **269**, 842–851 (2011).

48. Bennett, C. J., Pirim, C. & Orlando, T. M. Space-Weathering of Solar System Bodies: A Laboratory Perspective. *Chem. Rev.* **113**, 9086–9150 (2013).

49. Roussos, E. *et al.* Discovery of a transient radiation belt at Saturn. *Geophysical Research Letters* **35**, (2008).

50. Kollmann, P. *et al.* Spectra of Saturn's proton belts revealed. *Icarus* **376**, 114795 (2022).

51. Southworth, B. S., Kempf, S. & Spitale, J. Surface deposition of the Enceladus plume and the zenith angle of emissions. *Icarus* **319**, 33–42 (2019).

52. Moore, M. H. & Hudson, R. L. IR Detection of H2O2 at 80 K in Ion-Irradiated Laboratory Ices Relevant to Europa. *Icarus* **145**, 282–288 (2000).

53. Gomis, O., Satorre, M. A., Strazzulla, G. & Leto, G. Hydrogen peroxide formation by ion implantation in water ice and its relevance to the Galilean satellites. *Planetary and Space Science* **52**, 371–378 (2004).

54. Loeffler, M. J., Raut, U., Vidal, R. A., Baragiola, R. A. & Carlson, R. W. Synthesis of hydrogen peroxide in water ice by ion irradiation. *Icarus* **180**, 265–273 (2006).

55. Carlson, R. W. *et al.* Hydrogen Peroxide on the Surface of Europa. *Science* **283**, 2062–2064 (1999).

56. Trumbo, S. K., Brown, M. E. & Hand, K. P. H2O2 within Chaos Terrain on Europa's Leading Hemisphere. *AJ* **158**, 127 (2019).

57. Hendrix, A. R., Barth, C. A., Stewart, A. I. F., Hord, C. W. & Lane, A. L. Hydrogen Peroxide on the Icy Galilean Satellites. 2043 (1999).

58. Accolla, M. *et al.* Combined IR and XPS characterization of organic refractory residues obtained by ion irradiation of simple icy mixtures. *A&A* **620**, A123 (2018).

59. Baratta, G. A. *et al.* Organic samples produced by ion bombardment of ices for the EXPOSE-R2 mission on the International Space Station. *Planetary and Space Science* **118**, 211–220 (2015).

60. McCord, T. B. *et al.* Organics and Other Molecules in the Surfaces of Callisto and Ganymede. *Science* **278**, 271–275 (1997).